\documentclass[pre,aps,superscriptaddress,floatfix]{revtex4}

\usepackage{amsmath}
\usepackage{amssymb}
\usepackage{amsbsy}
\usepackage{graphicx}
\usepackage[usenames,dvipsnames,svgnames,table]{xcolor}

\usepackage{color}

\newcommand{\be}{\begin{equation}} 
\newcommand{\ee}{\end{equation}}
\newcommand{\bea}{\begin{eqnarray}}   
\newcommand{\eea}{\end{eqnarray}}

\begin{document}
 
\title{Self-propulsion against a moving membrane: enhanced accumulation and drag force}

\author{U. Marini Bettolo Marconi}
\affiliation{Scuola di Scienze e Tecnologie, 
Universit\`a di Camerino, Via Madonna delle Carceri, 62032, Camerino, INFN Perugia, Italy}

\author{A. Sarracino}
\affiliation{CNR-ISC and Dipartimento di Fisica, Sapienza Universit\`a di Roma, p.le A. Moro
2, 00185 Roma, Italy}

\author{C. Maggi}
\affiliation{CNR-NANOTEC and Dipartimento di Fisica, Sapienza Universit\`a di Roma, p.le A. Moro
2, 00185 Roma, Italy}

\author{A. Puglisi}
\affiliation{CNR-ISC and Dipartimento di Fisica, Sapienza Universit\`a di Roma, p.le A. Moro
2, 00185 Roma, Italy}

\begin{abstract}
Self-propulsion (SP) is a main feature of active particles (AP), such
as bacteria or biological micromotors, distinguishing them from
passive colloids. A renowned consequence of SP is accumulation at static interfaces, even in the absence of hydrodynamic
interactions. Here we address the role of SP in the interaction
between AP and a moving semipermeable membrane. In particular, we
implement a model of noninteracting AP in a channel crossed by a
partially penetrable wall, moving at a constant velocity $c$. With
respect to both the cases of passive colloids with $c>0$ and AP with
$c=0$, the AP with finite $c$ show enhancement of accumulation
in front of the obstacle and experience a largely increased drag
force. This effect is understood in terms of an effective potential
localised at the interface between particles and membrane, of height
proportional to $c\tau/\xi$, where $\tau$ is the AP's re-orientation
time and $\xi$ the width characterising the surface's smoothness ($\xi
\to 0$ for hard core obstacles). An approximate analytical
scheme is able to reproduce the observed density profiles and the
measured drag force, in very good agreement with numerical
simulations. The effects discussed here can be exploited for automatic
selection and filtering of AP with desired parameters.
\end{abstract}

\maketitle

\section{Introduction}

Active particles (AP) represent a large class of
systems characterized by a conversion of internal energy into self-propulsion~\cite{ram10}.  The behavior of AP
deeply differs from that of passive colloids in a thermal bath and
shows typical features of nonequilibrium dynamics~\cite{cates2012diffusive,BPM16}. At the level of
single trajectories, AP are characterized by persistent random walks
and correlated motion. Instances of such systems can be found in the
realm of bacteria and micro-organisms~\cite{marchetti2013}, or in the
context of man-made nano-devices~\cite{BLLRVV16}.

Several models have been proposed to study the physical properties of
active matter systems, which show intriguing phenomena, such as
nonequilibrium phase transitions, self-organization and collective
behaviors. Let us mention the ``run and tumble'' model~\cite{TC08},
characterized by directed motion interrupted by random reorientations,
the ``active Brownian'' model~\cite{G09,PCYB10}, where particles are
pushed by a constant force, whose direction changes stochastically,
and the Vicsek model~\cite{vicsek95,CGGR08}, where the particle speed
is fixed and the orientation depends on the average velocity of the
neighbors.  More recently, the Gaussian colored-noise (GCN) model has
been proposed to account for the correlated motion (over a typical
time $\tau$) characterizing AP systems~\cite{MBGL15}, which allows for
an analytical treatment within a specific scheme, known
as Unified Colored Noise Approximation (UCNA)~\cite{JH87}.

Among the several nonequilibrium phenomena observed in AP systems, a
surprising result reproduced also by the GCN model, is that, in the
presence of a static repulsive potential, AP do accumulate around the
obstacle, producing a nontrivial density
profile~\cite{wensink2008aggregation,geiseler2016}.  This observation raises the
question of what effects are produced when the obstacle is not static
and moves with constant velocity, inducing a
stationary current.

The study of the density profiles in (passive) colloidal systems under
the action of a moving obstacle, indeed, takes on great importance in
several contexts and has been addressed from different
perspectives. For instance, it is the central issue in active
microrheology, where a tracer is (magnetically or optically) driven
through a medium to probe its structural
properties~\cite{SqM09,PT14}. A moving potential barrier can also be
realized by means of optical fields, with travelling waves or inverted
traps~\cite{speckle1,speckle2,speckle3}. Moreover, soft potential barriers
with a finite height and width are also used to model the finite
thickness of a semipermeable membrane in contact with
fluids~\cite{bryk,marsh,margaritis,zwanzig1992diffusion}, or the
translocation properties of polymer chains through
nanopores~\cite{sung,ammenti}.
Similar problems related to the study of the stationary currents and
density profiles of colloids under the effect of moving potentials
have been addressed with the formalism of the density functional
theory, with applications to the motion of colloidal particles in
narrow channels~\cite{PT03}, or in polymer
solutions~\cite{PDT03,TM08}. 

\begin{figure}[!t]
\includegraphics[width=0.7\columnwidth,clip=true]{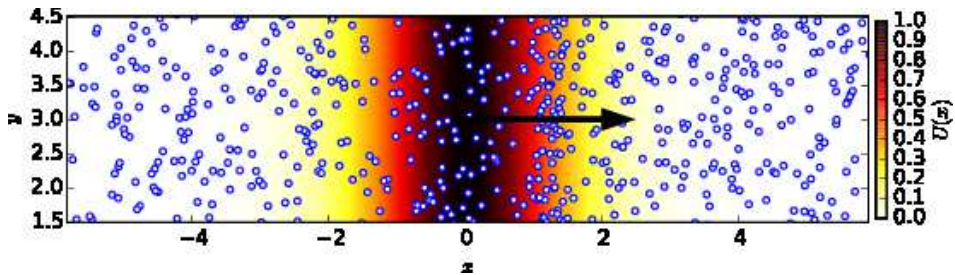}
\caption{A semipermeable membrane, modeled as a
  potential barrier $U(x)$ (color-bar), moves at velocity $c$ (denoted by the arrow) in a fluid of noninteracting active particles.}
\label{fig1}
\end{figure}

In this paper, we study a simple model for a semipermeable membrane
moving at constant velocity $c$ in a fluid of noninteracting GCN
active particles of persistence time $\tau$, see the sketch in
Fig.~\ref{fig1}.  Our analytical theory demonstrates the appearance of
an effective dynamical potential arising from the coupling of
self-propulsion with the nonequilibrium current induced by the moving
obstacle: indeed it vanishes in both the limits of $c\to 0$, and
$\tau\to 0$ (passive colloids with thermal noise).  Our approach,
which generalizes the UCNA to non-vanishing steady currents, gives
accurate predictions - when compared to numerical simulations - for
the density profiles of AP and the effective drag force, in a wide
range of parameters.  The most striking consequence of the
current-induced effective potential is an enhanced accumulation of AP
at the interface, with respect to the static case or with respect to
the behavior of passive colloids. This effect yields a drag force
whose intensity can be made large at will by tuning the model
parameters. In the nonlinear regime of large $c$, we also observe a
nonmonotonic behavior of the experienced drag
force~\cite{LF13,BIOSV14,BIOSV16}, which is well described within our
analytical approach.
Our results have practical applications, e.g. in
sweeping up AP from a mixture of inert/active particles, or in
selecting and filtering AP with specific parameters, by tuning the
properties of the moving membrane.

\section{Model}

 A channel, in generic dimension, contains suspended
(active or passive) particles. A membrane separates the channel in two
parts and moves with constant velocity $c$ along the direction $x$
perpendicular to itself, see Fig.~\ref{fig1}. Since the particles are
noninteracting, the only relevant direction is that parallel to the
membrane movement. We assume the channel to be periodic and very large
in the $x$ direction. The dynamics of each particle is described by
the overdamped Langevin equation
\begin{align} \label{adim}
  \dot{x}(t) &=\frac{F(x-c t)}{\zeta}+\eta(t), \\
   F(x) &= -d_x U(x),
\end{align}
where the potential $U(x)$ represents the moving penetrable membrane.
The width of the membrane is used as unit of length (see below,
Eq.~\eqref{eq:pot}), while the mass of the particle is 1. The quantity
$\eta(t)$ stands for a noise term, which is white (thermal) for passive
colloids, or coloured, with correlation time $\tau$, for active
particles: in both cases $\langle \eta(t) \rangle=0$. When
Eq.~\eqref{adim} models passive particles, we take $\langle \eta(t)
\eta(t')\rangle=\frac{2}{\zeta}\delta(t-t')$ and the host fluid has
unitary temperature: therefore $\zeta$ is the viscosity of the host
fluid in these particular units. When Eq.~\eqref{adim} models active
particles, $\eta(t)$ is GCN (``active noise''), i.e.
\begin{align}
  \dot{\eta}(t) &= -\zeta \eta(t) + \sqrt{2 \zeta} \chi(t)\\
  \langle\chi(t)\chi(t')\rangle &= \delta(t-t').
\end{align}
In this case $\zeta=1/\tau$ and the active effective temperature is
set to $1$ (or, equivalently, the active speed is set to $1$). We
notice that in both cases (passive and active), with chosen units, the
bare diffusion coefficient of the particles (i.e. when $U(x) \equiv
0$) is $1/\zeta$.  In the following, we use a smooth potential of the
form
\begin{equation} \label{eq:pot}
U(x) = U_0 \{\tanh[(x+1)/\xi] - \tanh[(x-1)/\xi]\},
\end{equation}
which is characterised by a steepness $1/\xi$.

In order to understand the main effects induced by self-propulsion
in the presence of a stationary current, we focus on two
quantities: i) the density profile around the moving obstacle
and ii) the experienced drag force.

\subsection{Effective potential}

 To proceed with our analysis, it is useful to notice that, when
$\eta(t)$ is GCN, we can time-derive Eq.~\eqref{adim}, obtaining
\begin{align}
  \dot{x}(t)&=v(t),
   \label{st1}
  \\
  \dot{v}(t)&=-\zeta g(x-c t) v(t) + F^*(x- ct) + \sqrt{2 \zeta} \chi(t),
  \label{st2}\\
  F^*(x) &= F(x)- \frac{c}{\zeta}\frac{d F(x)}{d x}= -\frac{d U(x)}{d x}+ \frac{c}{\zeta}\frac{d^2 U(x)}{d x^2},\\
  g(x) =&1+\frac{1}{\zeta^2}\frac{d^2 U(x)}{d x^2}.
  \end{align}
In the above equations two terms deserve discussion: an effective
force $F^*(x)$, which reduces to $-dU/dx$ when $c=0$, and an effective
viscosity $g(x)$. The latter - which is the only effect of
self-propulsion when $c=0$ - has been thoroughly discussed
in~\cite{MBGL15,MM15,MMM16}: it can be treated within an approximate
equilibrium-like solution (known as UCNA), based upon an effective
static potential $U_{stat}(x)=U(x)+\frac{1}{2 \zeta^2} \left( \frac{d
  U(x)}{d x} \right)^2- \ln |g(x)|$.  In the present case, the finite
velocity of the obstacle $c>0$ produces an additional contribution in
the force term, which is responsible for new dynamical effects. These
effects can be accounted for by a new approximate treatment  (see Appendix A).

\section{Dynamical UCNA}

 In the case of a shifting barrier, one rewrites the stochastic
differential equations \eqref{st1}-\eqref{st2} into the equivalent
Fokker-Planck equation for the probability distribution of position
and velocity $P(y,v)$:
\begin{eqnarray}
&&
\frac{\partial  P(y,v)}{\partial t} +v \frac{\partial }{\partial y}  P(y,v) 
+ F^*( y) \frac{\partial }{\partial v}  P(y,v) \nonumber\\
&&
=\zeta   \frac{\partial}{\partial v}\Bigl[
\frac{\partial }{\partial v }+g(y) v\Bigl]       P(y,v),
\label{kramers0b}
\end{eqnarray} 
with $y=x-ct$. In order to proceed further, we consider the steady state solution 
of Eq.~\eqref{kramers0b} and set $\frac{\partial  P(y,v)}{\partial t} =-c \frac{\partial  P(y,v)}{\partial y}$.   
By multiplying by powers of $v$ and integrating w.r.t.  $v$, one obtains a
hierarchy of coupled  first order ordinary differential equations for the velocity moments of $P(y,v)$, whose first two members are the continuity  equation for the density $\rho(y)=\int dv P(y,v) $ 
\begin{equation}
 -c \frac{d \rho (y)}{d y}+ \frac{d} {d y} J (y)=0,
\label{brinkman0}
\end{equation}
and the momentum balance equation for the current  $J(y)=\int dv v P(y,v) $:
\begin{equation}
-c\frac{d J(y)} {d y}  + \frac{d \Pi(y)} {d y} 
-F^*(y)\rho(y)-\zeta g(y) J(y) =0,
\label{brinkman1}
\end{equation} 
where $\Pi(y)=\int dv v^2 P(y,v)$.
According to Eq.~\eqref{brinkman0}
the current must be proportional to the density 
\begin{eqnarray}
J(y)=c [\rho(y)-\bar \rho],
\label{curcur}
\end{eqnarray}
where $\bar \rho$ is a constant such that the solution is periodic,  $\rho(L)=\rho(-L)$. 
The following distribution represents the exact solution of Eq.~\eqref{kramers0b} in the regions where the
force vanishes and contains adjustable parameters to obtain an approximate solution in the wall region: 
\begin{eqnarray}
P(y,v) &=& \sqrt{\frac{\beta(y)}{2\pi}} \left\{[\rho(y)-\bar \rho] \exp\left[ -\frac{1}{2} \beta(y) (v-c)^2\right] \right.\nonumber\\
 &+& \left.\bar \rho \exp\left[ -\frac{1}{2} \beta(y) v^2\right]\right\},
 \label{distribution}
\end{eqnarray}
where $\beta(y)$ is a positive definite function.  Remarkably,
expression~\eqref{distribution} also represents an (approximate)
closure of the infinite hierarchy of equations (of which
Eqs.~\eqref{brinkman0} and~\eqref{brinkman1} are the first two
members) generated by the transformation of the partial differential
equation ~\eqref{kramers0b} into a set of coupled ordinary
differential equations for the velocity moments of $P$.  Hence,
according to the information contained in Eq.~\eqref{distribution} the
momentum flux reads $\Pi(y)= \frac{\rho(y)}{\beta(y)}+c^2 [
  \rho(y)-\bar \rho]$, so that Eq.~\eqref{brinkman1} becomes:
\begin{equation}
\frac{d } {d y}  \frac{\rho(y)}{\beta(y)}
=[F(y)-\zeta c ]\rho(y)+\zeta c g(y)\bar \rho, 
\end{equation}
which has the following interpretation: the ``active pressure'' gradient $\frac{d}{d y} \frac{\rho(y)}{\beta(y)}$  is balanced by the 
force due to the moving wall and by the friction force $-\zeta g(y) J(y)$ (the second term in the r.h.s.).
In the
case of a very weak potential,  $\rho(y)\approx \bar \rho$ and
the current vanishes, whereas for high barriers
$\rho(y)\gg\bar \rho$ and $J(y)\approx c\rho(y)$.
The static UCNA approximation is recovered by setting $c=0$, i.e. $J=0$ and $\beta(y)=g(y)$. The density profile is given by:
\begin{eqnarray}
 \frac{\rho(y)}{\beta(y)}&=&\frac{ \rho(L)}{\beta(L)}  e^{-w(y)+w(-L)-c\zeta (y+L) }\nonumber\\
 &+& \zeta c \bar \rho  e^{-w(y)-\zeta c y}  \int_{-L}^{y} ds e^{w(s)+c\zeta s} g(s),
\end{eqnarray}
where $w(y)$ is an effective potential defined by:
\begin{equation}
w(y)=\int_{-L}^y ds \beta(s) \frac{d U(s)}{d s}+ \zeta c \int_{-L}^y ds [\beta(s)-1],\label{wy}
\end{equation}
and $\rho(L)$ is fixed by the normalization of the number of
particles. The explicit expression of the constant $\bar \rho$ is
given in Appendix A.  Interestingly, the second term in the r.h.s. of
Eq.~(\ref{wy}) can be identified with a dynamical potential
$U_{dyn}(y)$ vanishing when either $c=0$ (static barrier) or $\zeta\to
\infty$ (passive particles). Therefore it is a peculiar feature of our
model, arising from the coupling of self-propulsion with the
nonequilibrium current.  This term gives an effective trap -- at the
front of the moving potential -- of height $\sim U_0 c/(\zeta \xi)$,
and a specular effective barrier at its tail.  As one can see, the
solution for $c\neq 0$ is not Boltzmann-like since the system is in a
truly nonequilibrium state and therefore the density profile is not
symmetric with respect to the transformation $y \to -y$ which
characterizes the bare potential $U(y)$.

Since the UCNA breaks down in regions with negative curvature of the
potential~\cite{JH87}, for the purpose of obtaining quantitative predictions for
$\rho(y)$, we empirically set $\beta(y)=g(y)$ where $g(x)\ge 0$, and
$\beta(y)=0$ otherwise. From the density profile $\rho(y)$ we also obtain the average drag
force acting on the moving barrier $\langle F\rangle =\int_{-L}^L dy
F(y)\rho(y)$, which obeys the sum rule
\begin{equation} \label{drag}
\langle F\rangle  = \zeta c \int_{-L}^L dy [\rho(y)- g(y)\bar \rho].
\end{equation}

\section{Numerical results}

 The approximations underlying our theory
have been fairly verified by comparison with numerical simulations of
the model in Eq.~\eqref{adim}, for both passive and active
particles. The simulations implement a time-discretized scheme for
Eq.~\eqref{adim} through a fourth-order Runge-Kutta
algorithm~\cite{H92}, with a time step $dt=10^{-4}$. Averages are done
on a single trajectory of length $5 \times 10^8$ in the used units. In
the figures, error bars fall within the symbols.

\begin{figure}[!t]
\includegraphics[width=0.6\columnwidth,clip=true]{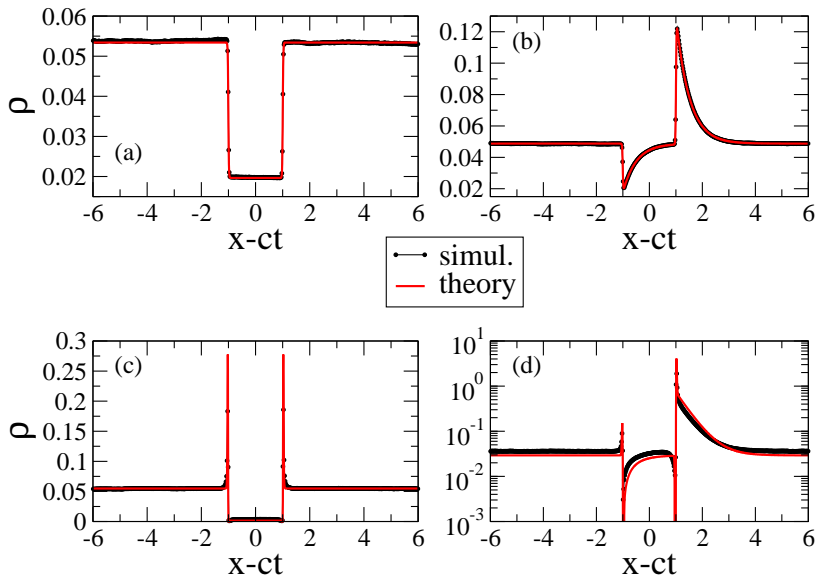}
\caption{Density profiles with different kinds of noise (thermal or active) and different values of the barrier's velocity $c$: (a) thermal noise, $c=0$: (b) thermal noise, $c=0.2$; (c) active noise, $c=0$; (d) active noise, $c=0.2$.}
\label{fig2}
\end{figure}

In Fig.~\ref{fig2} we show the density profiles for two passive cases
and two active cases, with static or moving potential. A first
important information is the good match between simulations and
theory. In the passive case (two top frames), switching on the
external velocity from $c=0$ to $c>0$ leads to an imbalance of the
density distribution with an accumulation at the front of the membrane
(at $x-ct=1$, see expression of the moving potential,
Eq.~\eqref{eq:pot}), and a depletion at its tail (at
$x-ct=-1$)~\footnote{Our theory predicts an asymptotic exponential
  decay of the density profiles both in front and past the moving
  wall, in agreement with what found for analogous problems in lattice
  systems~\cite{BIOSV16}.}.  The two frames on the left ($c=0$)
demonstrate that switching from passive to active particles induces an
accumulation of particles near both borders of the membrane potential,
with a depletion inside the energetically unfavoured region. The novel
effect discussed here appears, strikingly, in the active case with
$c>0$ (bottom-right frame): the accumulation of particles on the
moving front of the membrane becomes much more important than the
passive case with $c>0$ or the active case with $c=0$ (notice the log
scale on the $y$ axis).

\begin{figure}[!t]
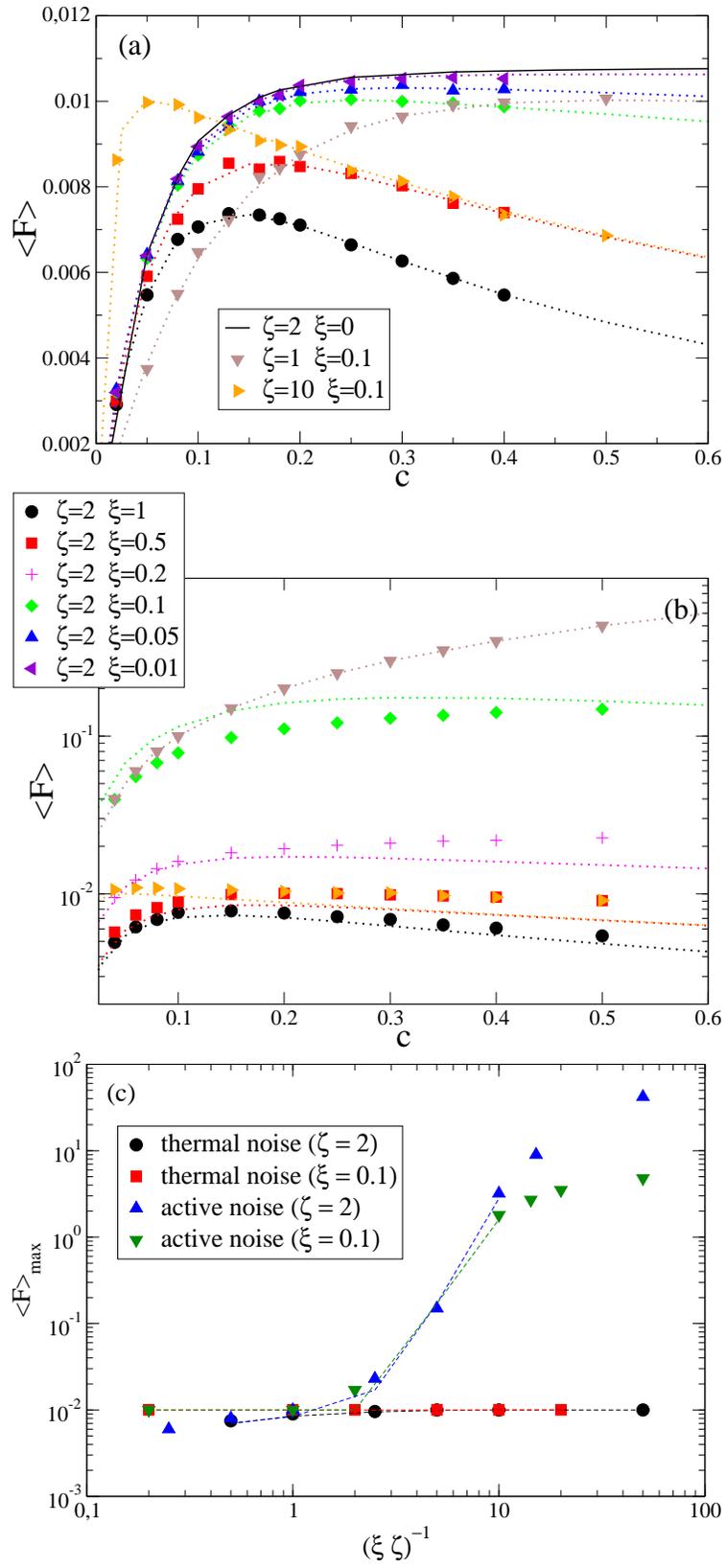

  \includegraphics[width=0.55\columnwidth,clip=true]{fig3anew.eps}\\
\includegraphics[width=0.55\columnwidth,clip=true]{fig3bnew.eps}\\
\includegraphics[width=0.55\columnwidth,clip=true]{max.eps}
  \caption{(a) and (b): average drag force versus velocity $c$ for thermal noise (a) and active noise (b).  Symbols
    represent numerical simulations (legends are valid for both
    graphs), while dotted lines are the theoretical predictions; (c):
    maximum average drag force versus $(\xi \zeta)^{-1}$. Symbols are
    numerical simulations, while lines are from theory.}
\label{fig3}
\end{figure}

To understand the behavior of the system in full generality, exploring
the effects of all parameters, we focus on the global observable
$\langle F(y) \rangle$, which is the average drag force
experienced by the moving membrane. Several results are shown in
Fig.~\ref{fig3}A and~\ref{fig3}B, where a comparison is presented between passive
and active cases for several values of
$\zeta$ and $\xi$ in a relevant range of velocities $c$. Again, we
observe a fair superposition of numerical results with theoretical
predictions, Eq.~\eqref{drag}: this is expected for the passive cases,
where the theory is exact, while it is not trivial at all in the
active case.  Surprisingly, even at low $\xi$, a
reduction of $\zeta$ (longer activity persistence time $\tau$)
may improve the agreement with the simulations.


In all the cases considered (excluding the $\xi \to 0$ limit for the
passive case), at constant $\xi$ and $\zeta$, the average drag reaches
a maximum at some value $c^*$ and then decreases for $c > c^*$. This
can be understood in terms of competition between ``kinetic energy''
$\sim c^2$ and the potential barrier. In the passive case this leads
to a value of $c^*$ which is roughly independent of $\xi$ or $\zeta$,
and a saturation of $\langle F \rangle$ when $\xi \to 0$, as seen in
Fig.~\ref{fig3}A. In the active case at large $1/(\xi \zeta)$ the
dynamic potential $\sim U_0 c/(\zeta\xi)$ dominates, so that the
energetic argument leads to $c^* \sim U_0/(\zeta\xi)$. When the
effective barrier is high and $c < c^*$, very few particles cross it
and the majority goes at $\dot{x} \sim c$, so that Eq.~\eqref{adim} on
average gives the linear behavior $\langle F \rangle \approx \zeta c$,
well visible in simulations at large values of $1/(\zeta
\xi)$. Estimating the maximum value of the drag force to be $\langle F
\rangle_{max} \approx \zeta c^*$, we get for the active case $\langle
F \rangle_{max} \approx U_0/\xi$, expected to hold at large $(\zeta
\xi)^{-1}$. The active case with a moving membrane, therefore, is {\em
  qualitatively} different from the passive case - or from any case at
$c=0$ - since the average drag force can increase indefinitely by
reducing $\xi$. In Fig.~\ref{fig3}C we have shown $\langle F
\rangle_{max}$ versus $(\zeta \xi)^{-1}$ for the active and passive
cases: at intermediate values of $(\zeta \xi)^{-1}$ an interesting
data collapse is found, together with a sharp increase with $(\zeta
\xi)^{-1}$ for the active case. Such an increase eventually saturates
if $\zeta$ is further decreased at constant $\xi$, or continues if
$\xi$ is reduced at constant $\zeta$, demonstrating the qualitative
difference between the active and the passive cases.

\section{Conclusions}

 We have shown the existence of a dynamical
enhancement of clustering and drag when a travelling barrier sweeps
active particles. The synergy of two dynamical effects (active noise
and non-zero current) leads to a scenario qualitatively new,
as shown in Fig.~\ref{fig3}C: indeed the average drag is sensitive to
the persistence time $1/\zeta$ and to the steepness of the membrane
potential $1/\xi$, and can be made indefinitely strong. We have discussed
a theoretical treatment of this effect, fairly compared with numerical
simulations. This is remarkable if one considers that predictive
theoretical schemes are scarce in the framework of active
particles, particularly in the non-linear regime with strong spatial currents as in our case.
It is  interesting to note that our theory
truncates the Fokker-Planck hierarchy at the same order of the static
UCNA scheme: however, unlike the static UCNA, it leads to a genuine
{\em non-equilibrium} behavior~\cite{FNCTVW16,BPM16}.



The parameter values used in our simulations are in the range of
realistic systems of AP, therefore they are within reach for
experimental verification, e.g. in setups with optical travelling
waves or inverted traps~\cite{speckle1,speckle2,speckle3}, taking care
to avoid competing effects such as diffusiophoretic
torques or hydrodynamic-induced
wall-attachment~\cite{wysocki15,lozano16}. For instance, taking as
unit of length $10 \mu m$ (order of magnitude of the width of
lithographed micro-membrane), typical biological swimmers with speed
$\sim 10 \mu m/s$ and reorientation time $\tau \sim 1 s$ correspond to
$\zeta \approx 1$. A straightforward application of our study is the
possibility to separate a mixture of AP, filtering out those with
given parameters (e.g. a certain value of $\tau$) by sweeping a
membrane with well-tuned values of $c$ and $\xi$.

\appendix

\section{Model equations}
We consider a dilute solution of active particles dragged along the
x-direction under the action of a travelling potential barrier with
velocity $c$, modelled by a time dependent external potential,
$U(x,t)=U(x-ct)$, which acts on the colloidal particles but has
negligible effects on the solvent
\cite{penna2003dynamic,penna2003dynamicb,tarazona2008beyond}.  For the
sake of simplicity we neglect the interactions among the particles and
any hydrodynamic effect and include only the friction, through a drag
coefficien $\gamma$.  The active forces are modelled by a coloured
noise, i.e. Gaussian noise with exponential memory of characteristic
time $\tau$. Note that, in this Appendix, we introduce
the model with all dimensional parameters and explicitly show the
change of variables necessary to obtain the Equations studied in the
paper.

\subsubsection{Langevin description} 
 The following stochastic dynamics is  assumed
\begin{equation}
\label{uno}
\dot{x}(t) =\frac{1}{\gamma} F(x,t) + \eta(t) \, ,
\end{equation}
where $F= -\partial U/\partial x$ and $\eta$ mimics the self-propulsion mechanism and is assimilated to an Ornstein-Uhlenbeck process 
\begin{equation}
\dot \eta(t) =- \frac{1}{\tau} \eta(t) + \frac{ D^{1/2}}{ \tau} \xi (t)  .
\label{due}
\end{equation}
The underlying stochastic force $\xi(t)$ is a Gaussian and Markovian process distributed with zero mean
and moments  $\langle \xi(t) \xi(t')\rangle=2    \delta(t-t')$.  
The coefficient $D$ due to the activity is related to the correlation of the 
Ornstein-Uhlenbeck process  $\eta(t)$ via 
\be
\langle \eta(t) \eta(t')\rangle=\frac{D}{\tau} \exp \Bigl( -\frac{|t-t'|}{\tau} \Bigr) .
\ee
\subsubsection{Fokker-Planck description} 
After differentiating  with respect to time eq.~\eqref{uno} and introducing a velocity $v=\dot x$, we may write the following system of equations:
\bea
\dot x&=&v
 \nonumber\\
 \dot v &=&-\frac{1}{\tau}\left(1-\frac{\tau}{\gamma} \frac{\partial F}{\partial x}\right) v+ \frac{1}{\tau\gamma} \Bigl( F+\tau \frac{\partial F}{\partial t} \Bigr) +\frac{ D^{1/2}}{ \tau} \eta \, .
 \eea
 The latter equation in the case of the shifting potential becomes:
 \be
 \dot v=-\frac{1}{\tau}\left(1-\frac{\tau}{\gamma} \frac{\partial F}{\partial x}\right) v+ \frac{1}{\tau\gamma}\Bigl( F-\tau c\frac{\partial F}{\partial x}\Bigr) +\frac{ D^{1/2}}{ \tau} \eta
\label{sistema2}
\ee
and the associated Fokker-Planck (FP) equation for the "phase-space" distribution $P(x,v,t)$ reads
\be
\frac{ \partial P }{\partial t}+ v \frac{\partial P}{\partial x}+\Bigl(\frac{F(x,t)+ \tau  \frac{\partial F(x,t)}{\partial t}  }{ \gamma \tau}\Bigr) \frac{\partial P}{\partial v}
= \frac{1}{ \tau}  \frac{\partial }{\partial v} \Bigl(  \frac{D}{\tau}\frac{\partial }{\partial v}+    g(x,t) v \Bigl) P \, ,
\label{kunobis}
\ee
where $g(x,t)=\left(1-\frac{\tau}{\gamma} \frac{\partial F}{\partial x}\right)$.

Now, defining $F^*=(F-\tau c\frac{\partial F}{\partial x})$ and
considering the steady state regime of the system, where $P(x,v,t)$ must have the travelling wave  form $P(x-ct,v)$, we can write:
 \be
-c\frac{ \partial P }{\partial x}+ v \frac{\partial P}{\partial x}+\frac{F^*(x-ct)}{ \gamma \tau} \frac{\partial P}{\partial v}
= \frac{1}{ \tau}  \frac{\partial }{\partial v} \Bigl(  \frac{D}{\tau}\frac{\partial }{\partial v}+    g(x-ct) v \Bigl) P .
\label{kunob}
\ee 
In the problem at hand,
the shifting external potential $U$  is localized within a finite region around the origin of the  comoving reference frame 
and vanishes for $x\to \pm \infty$.
\subsubsection{Non dimensional  variables}
In order to proceed further, it is time saving to adopt non dimensional variables for positions, velocities,  and time,
and rescale forces accordingly.
We define $v_T=\sqrt{D/\tau}$, measure lenghts using the characteristic lenght, $\ell$, of the potential  and  introduce 
the following non dimensional variables: 
\begin{equation}
\bar t\equiv t\frac{v_T}{l}, \qquad \bar v\equiv\frac{v}{v_T}, 
\qquad X\equiv\frac{x}{\ell},\qquad   \bar F(\bar x,\bar t)\equiv\frac{\ell F(x,t)}{ D\gamma}    ,\qquad \zeta=\frac{\ell}{\tau v_T} ,\qquad  \bar P= v_T \ell\, P ,\qquad \bar c=\frac{c}{v_T} ,
\label{adim1}
\end{equation}
where $\zeta$ plays the role of a non dimensional friction. To lighten the notation we shall drop the bar over the non dimensional 
variables without incurring in ambiguities.

In the case of a shifting barrier, one can write the following Fokker-Planck equation in terms of the coordinate
$y=x-ct$ relative to the comoving reference frame:
\begin{eqnarray}
-c \frac{\partial  P(y,v)}{\partial y} +v \frac{\partial }{\partial y}  P(y,v) 
+ F^*( y) \frac{\partial }{\partial v}  P(y,v) 
=\zeta   \frac{\partial}{\partial v}\Bigl[
\frac{\partial }{\partial v }+g(y) v\Bigl]       P(y,v) \, 
\label{kramers0bs}
\end{eqnarray} 
\subsubsection{Hydrodynamic theory}
 In order to proceed further, it is convenient to eliminate the $v$ dependence of the 
phase-space distribution $P(y,v)$, by multiplying by powers of $v$ and integrating w.r.t.  $v$. One obtains a
set of coupled  first order ordinary differential equations, the so-called Brinkman hierarchy, whose first two members are the continuity  equation and the momentum balance equation, respectively:
\begin{eqnarray}
 &&
 -c \frac{d \rho (y)}{d y}+ \frac{d} {d y} J (y)=0 \, ,
\label{brinkman0s}
\\
&&
-c\frac{d J(y)} {d y}  + \frac{d \Pi(y)} {d y} 
-F^*(y)\rho(y)+\zeta g(y) J(y) =0
\label{brinkman1s}
\end{eqnarray}
where we have introduced the density $\rho(y)$, the current $J(y)$ and the momentum current $\Pi(y)$, respectively, via:
\bea
\rho(y)&=&\int dv P(y,v)  \, ,\\
J(y)&=&\int dv v P(y,v) \, , \\
\Pi(y)&=&\int dv v^2 P(y,v)  \, .
\eea
According to the continuity equation \eqref{brinkman0s}
the current must be proportional to the density 
\begin{eqnarray}
J(y)=c [\rho(y)-\bar \rho] \, ,
\label{curcurs}
\end{eqnarray}
where $\bar \rho$ is a constant such that the solution is periodic at $\rho(L)=\rho(-L)$, where $2 L$ is the box size.
 As we shall see later, for  large systems $L \gg \ell$, $\bar \rho \approx \rho(\pm L)$
and the current is almost vanishing at the boundaries.

It can be easily verified that the following distribution is a solution of the eq. \eqref{kramers0bs}
in regions where $F^*(y)=0$ and $g(y)=1$:
\be
P(y,v)=  \Bigl[\rho(y)-\bar \rho\Bigr] \, H_0(v-c)+ \bar \rho \, H_0(v) \, ,
\label{doublepeak0} 
\ee
where 
\be
H_0(v)=\frac{1}{\sqrt{2\pi}}  \exp \Bigl(-\frac{1}{2}  v^2 \Bigr)
\ee
is a Hermite function of zeroth order. By substituting  the ansatz~\eqref{doublepeak0} in eq.~\eqref{kramers0bs} (with $F=0$)
we obtain a solution provided $\rho(y)$ satisfies the following
condition:
\be
\frac{d  \rho(y)}{d y}= - \zeta c  (\rho(y)-\bar \rho) \, .
\ee

\subsubsection{Solution in the presence of a force field}

Now, we insist in looking for a solution of eq. ~\eqref{kramers0bs} of the form:
\be
P(y,v)=  \Bigl[\rho(y)-\bar \rho\Bigr] \, H_0(y,v-c)+ \bar \rho \, H_0(y,v) \, ,
\label{doublepeak} 
\ee
even in the region where $F(y)\neq 0$.  
We have introduced  the following (non uniform) Hermite functions, which are position dependent through $\beta(y)$, an adjustable function:
\bea
 H_0(y,v)&=& \sqrt{\frac{\beta(y)}{2\pi}}   \exp \Bigl( -\frac{\beta(y)}{2}  v^2 \Bigr) \, , \\
 H_1(y,v) &=& \sqrt{\frac{\beta(y)}{2\pi}}  \beta^{1/2}(y)\, v\, \exp \Bigl( -\frac{\beta(y)}{2}  v^2 \Bigr)  \, .
\eea
If we do that, i.e. if we apply the full FP operator to the trial distribution~\eqref{doublepeak}  we get:
\bea
&& H_1(y,v-c)\frac{1}{\beta^{1/2}} \Bigl(   \frac{d  \rho(y)}{d y} -\beta (F^*-\zeta g c) (\rho -\bar \rho) -\frac{\beta'}{\beta}(\rho-\bar\rho) \Bigr)  -H_1(y,v) )\frac{1}{\beta^{1/2}}  \Bigl( \beta F^* \bar \rho +\frac{\beta'}{\beta} \bar \rho \Bigr)\nonumber\\
&&+\zeta (g-\beta)\Bigl((\rho-\bar \rho) H_2(y,v-c)+\bar \rho H_2(y,v)\Bigr)
-\frac{\beta'}{2 \beta^{3/2}} \Bigl( (\rho -\bar \rho) H_3(y,v-c)+\bar \rho H_3(y,v)-c \beta^{1/2} H_2(y,v)\Bigr)=0 \, , \nonumber\\
\label{failedattempt}
\eea
where $H_2(y,v)$ and $H_3(y,v)$ are the Hermite functions of order 2 and 3, respectively,
and given by the recursion relation:
 $$H_{\nu+1}(y,v)=-\frac{1}{\beta^{1/2}}\frac{\partial H_\nu(y,v)}{\partial v}.$$
 The trial solution fails to solve eq. ~\eqref{kramers0bs}. However, if we limit ourselves to consider only the two lowest moments of the probability distribution, i.e. if   after multiplying by $(v-c)$, we integrate \eqref{failedattempt} over $v$ we obtain the following condition which gives the profile equation:
\be
  \frac{1}{\beta} \frac{d  \rho(y)}{d y} -(F-\zeta c) \rho(y)  -\frac{\beta'}{\beta^2}\rho(y) -\zeta c g \bar \rho =0 \, .
\label{momentouno}
\ee
If we continue the projection procedure beyond the first order in $(v-c)$ there will be an error in the equation for the second moment, which becomes inconsistent with the value of the second moment imposed by the trial distribution
(which, in fact, is already fixed by the trial form and therefore does not contain enough parameters to satisfy the extra conditions.).

\subsubsection{Construction of the solution}
Eq.~\eqref{momentouno} can be rearranged as follows:
\begin{equation}
\frac{d } {d y} \frac{\rho(y)}{\beta (y)}
=(F(y)-\zeta c )\rho(y)+\zeta c g(y)\bar \rho .
\label{mainequation}
\end{equation}

Notice that  the  ansatz for the phase-space distribution, gives the following expression for the  momentum flux:
\begin{equation}
\Pi(y)= \frac{ \rho(y)}{\beta (y)}+c^2 ( \rho(y)-\bar \rho) \, .
\label{momentumflux}
\end{equation} 
Notice that eq.~\eqref{momentouno} is perfectly equivalent to eq.~\eqref{brinkman1s} when the latter is endowed with a closure,
indeed represented by eq.~\eqref{momentumflux}.
The static UCNA approximation is recovered by setting the arbitrary function $\beta (y)=g(y)$ and  $c=0$, (i.e. $J=0$).
The solution of the inhomogeneous equation in the case of $c\neq 0$ is
\be
 \rho(y)=\frac{ \rho(L)}{\beta(L)} \beta(y)  e^{-(w(y)-w(-L))-c\zeta (y+L) }
 \Bigl[1+ (e^{2\zeta c L}  e^{w(L)-w(-L)} -1)\frac{ \int_{-L}^{y} ds e^{w(s)+c\zeta s} g(s)}{ \int_{-L}^{L} ds e^{w(s)+c\zeta s} g^(s)   }     \Bigr] \, ,
\ee
where $\rho(L)$ is fixed by the normalization and
the effective potential $w(y)$  is  defined by
\be
w(y)=\int_{-L}^y ds \beta(s) \frac{d U(s)}{d s}+ \zeta c \int_{-L}^y ds [\beta(s)-1] \, ,
\ee
The function $\beta(y)$ is given by $g(y)$ when $g(y)>0$ and  $\beta(y)=0$ otherwise.

\subsubsection{ Average Force and sum rule}

The average drag force is given by
\be
<F>=\int_{-L}^L dy F(y)\rho(y)=\zeta c \int_{-L}^L dy [\rho(y)-\bar \rho g(y)].
\ee
The constant $\bar \rho$ is
$$
\bar \rho=\frac{1}{\zeta c}\frac{\rho(L)} {\beta(L)}   \frac{  \Bigl(e^{ w(L)+c\zeta L}-e^{ w(-L)-c\zeta L} \Bigr )  }{ \int_{-L}^L dy e^{w(y)+c\zeta y} g(y) } 
$$
and we can rewrite the solution as:
\begin{eqnarray}
 \frac{\rho(y)}{\beta(y)}=\frac{ \rho(L)}{\beta(L)}  e^{-w(y)+w(-L)-c\zeta (y+L) }+ \zeta c \bar \rho  e^{-w(y)-\zeta c y}  \int_{-L}^{y} ds e^{w(s)+c\zeta s} g(s).
\end{eqnarray}

Finally, in order to regularize the problem we have chosen $\beta(y)=g(y)$ when $g(y)>0$ and $\beta(y)=0$ otherwise.


\section{The Dual picture}
The same mathematical problem can describe a different physical set up.
Consider a one dimensional system and a non uniform potential $U(y)$ acting in a central region only, where $F(y)=-\frac{dU}{dy}\neq 0$.
The particles are subject to colored noise and to a uniform force $E$. There will be a constant current, say $J_0$.

The obstacle is fixed in space, represented by the force $F(y)$.
There is a constant external field $E$
\be \label{dual}
\dot y=\frac{F(y)+E}{\gamma}+\eta(t)
\ee
where $\eta(t)$ is the standard colored noise as before. Time-differentiating Eq.~\eqref{dual} we get
\bea
&& \dot y= v \\
&& \dot v=\frac{F'(y)}{\gamma} v -\frac{v(t)}{\tau}+ \frac{F+E}{\tau \gamma}    +\frac{D^{1/2}}{\tau}\xi \, .
\eea
Equivalently, we write the associated FP equation:
\be
\frac{ \partial P }{\partial t}+ v \frac{\partial P}{\partial y}+\Bigl(\frac{F(y) +E }{ \gamma \tau} \Bigr)\frac{\partial P}{\partial v}
= \frac{1}{ \tau}  \frac{\partial }{\partial v} \Bigl(  \frac{D}{\tau}\frac{\partial }{\partial v}+    g(y) v \Bigl) P \, ,
\label{kunobis0}
\ee
which for a stationary system becomes
\be
 v \frac{\partial P}{\partial y}+\Bigl(\frac{F(y) +E }{ \gamma \tau}\Bigr) \frac{\partial P}{\partial v}
= \frac{1}{ \tau}  \frac{\partial }{\partial v} \Bigl(  \frac{D}{\tau}\frac{\partial }{\partial v}+    g(y) v \Bigl) P \, .
\label{kunobis2}
\ee
In non dimensional form we have:
\be
v \frac{\partial }{\partial y}  P(y,v) 
+ (F( y)+E) \frac{\partial }{\partial v}  P(y,v) 
=\zeta   \frac{\partial}{\partial v}\Bigl[
\frac{\partial }{\partial v }+g(y) v\Bigl]       P(y,v) \,.
\label{kramers0nb}
\ee
If one integrates over $v$ and defines $J(y)=\int dv v P(y,v)$, one finds:
\be
\frac{d }{d y}  J(y) =0 \, .
\ee
 The current is, now, constant: $J(y)=J_0$.
Let us multiply by $v$ and integrate eq. \eqref{kunobis2}:
\be
 \frac{d }{d y}  \Pi(y)
- (F( y)+E)  \rho(y)
=-\zeta   g(y)  J_0 \, ,
\label{kramerspb}
\ee
with $\Pi(y)=\int v^2 P(y,v) dv$.
Let us invert the relation and make the ansatz:
\be
P(y,v) =\bar \rho H_0(y,v-u)+(\rho(y)-\bar \rho) H_0(y,v)
\label{ansatzh}
\ee
Substituting \eqref{ansatzh}  in eq. \eqref{kramers0nb} when $F=0$ and $g=1$ we obtain
\be
H_1(v) \Bigl(\frac{d \rho(y)}{d y} -E(\rho(y)-\bar \rho) \Bigr)
-\Bigl( E \bar \rho-\zeta c \bar \rho \Bigr) H_1(v-u) =0 \, ,
\ee
whose solution is:
\bea
&&
\frac{d \rho(y)}{d y} -E(\rho(y)-\bar \rho) =0\\
&&
E\bar \rho=\zeta u \bar \rho
\eea

Now, we go back to eq. \eqref{kramerspb} and use the following closure (contained already in the parametric form of the solution for $P(y,v)$):
\bea
J(y)&=&J_0=\bar \rho u \, , \\ 
\Pi(y)&=&\bar \rho u^2+\frac{\rho(y)}{\beta(y)} \, .
\eea
So that the equation for $\rho(y)$ reads:
\be
 \frac{d }{d y}  \frac{\rho(y)}{\beta(y)}
- (F( y)+E)  \rho(y)
=-\zeta  u  g(y)  \bar \rho \, .
\label{kramersqb}
\ee
Now, such an equation is identical to the equation \eqref{mainequation} , provided we identify:
\bea
E&=&-\zeta c \,  ,\\
u&=&-c \, , \\
u&=&\frac{E}{\zeta} \, .
\eea
Thus, we  have shown that the equation for $\rho(y)$ is of the same type as the Nernst-Planck  (NP) equation:
The NP equation assumes that the constant current $J_0$ results from the combined effects of 
a diffusive current due to the random fluctuations ( the "thermal agitation" in other words)  and  a deterministic migration current
due to the coupling  to an external field $E$, which can be also modified by the presence of some localized potential
$U=-\int^y F(s) ds$:
\be
J_0= -\frac{1}{\zeta}  \frac{1} {g(y)} \frac{d }{d y}  \frac{\rho(y)}{\beta(y)}
+\frac{1}{\zeta g(y) } (F( y)+E)  \rho(y) \, ,
\label{kramersqbu}
\ee
with a space-dependent  diffusion coefficient 
\be
D(y)\equiv \frac{1}{\zeta}  \frac{1} {\beta(y) g(y)} 
\ee
and a space-dependent mobility
\be
\mu(y)=\frac{1}{\zeta}  \frac{1} {g(y)} \, .
\ee
Notice that this is exactly the UCNA equation for the current, which can be derived
without phase-space considerations.

Finally, let us rewrite
\be
J_0= - \frac{d }{d y} \Bigl ( D(y) \rho(y)\Bigr)
+\mu(y)  \Bigl( F( y)+E -  \frac{1}{\beta(y)} \frac{d}{d y} \ln g(y) \Bigr)  \rho(y)
\ee
There is an extra contribution from the drift stemming from the colored noise.

Note that the mathematics is the same as for the original problem, but
the interpretation of each term is now different.  If we look at the
profiles, we observe a crowding of active particles at the front of
the potential (where the derivative of $U$ is largest) and a depletion
inside. Particle near the entrance loose mobility and therefore crowd
there.
With strong activity and sharp entrances
($\zeta \to 0$ and $\xi\to 0$, respectively) the current should go to
zero.

\bibliography{biblio}

\end{document}